\documentclass[aps,prl,twocolumn,showpacs,superscriptaddress]{revtex4-1}  
\usepackage{amsmath}
\usepackage{amssymb}
\usepackage{graphicx}
\usepackage{epstopdf}
\usepackage{gensymb}
\usepackage[dvipsnames]{xcolor}
\usepackage{multirow}
\usepackage{hhline}

\renewcommand{\vec}{\mathbf}

\definecolor{HH-color}{named}{magenta}
\definecolor{HH-color2}{named}{NavyBlue}

\begin{document}

\setlength{\abovedisplayskip}{3.5pt}
\setlength{\belowdisplayskip}{3.5pt}

\title{Non-Hermitian topology of one-dimensional spin-torque oscillator arrays}

\author{Benedetta Flebus}
\address{Department of Physics, The University of Texas at Austin, Austin, TX 78712, USA}
\author{Rembert A. Duine}
\address{Institute for Theoretical Physics and Center for Extreme Matter and Emergent Phenomena,
Utrecht University, Leuvenlaan 4, 3584 CE Utrecht, The Netherlands}
\address{Department of Applied Physics, Eindhoven University of Technology,
P.O. Box 513, 5600 MB Eindhoven, The Netherlands}
\address{Center for Quantum Spintronics, Department of Physics,
Norwegian University of Science and Technology, NO-7491 Trondheim, Norway}
\author{Hilary M. Hurst}
\address{Joint Quantum Institute, National Institute of Standards and Technology,
and University of Maryland, Gaithersburg, Maryland, 20899, USA}
\address{Department of Physics and Astronomy, San Jos\'e State University, San Jos\'e, California, 95192, USA}

\begin{abstract}
Magnetic systems have been extensively studied both from a fundamental physics perspective and as building blocks for a variety of applications. Their topological properties, in particular those of excitations, remain relatively unexplored due to their inherently dissipative nature. The recent introduction of non-Hermitian topological classifications opens up new opportunities for engineering topological phases in dissipative systems. Here, we propose a magnonic realization of a non-Hermitian topological system. A crucial ingredient of our proposal is the injection of spin current into the magnetic system, which alters and can even change the sign of terms describing dissipation. We show that the magnetic dynamics of an array of spin-torque oscillators can be mapped onto a non-Hermitian Su-Schrieffer-Heeger model exhibiting topologically protected edge states. Using exact diagonalization of the linearized dynamics and numerical solutions of the nonlinear equations of motion, we find that a topological magnonic phase can be accessed by tuning the spin current injected into the array. In the topologically nontrivial regime, a single spin-torque oscillator on the edge of the array is driven into auto-oscillation and emits a microwave signal, while the bulk oscillators remain inactive. Our findings have practical utility for memory devices and spintronics neural networks relying on spin-torque oscillators as constituent units. 
\end{abstract}



\maketitle

\textit{Introduction}.~Since the discovery of the quantum Hall effect in 1980, topology has become a cornerstone in the understanding of condensed matter systems~\cite{topo}. Connecting concepts drawn from topology to electronic systems has led to the discovery of a plethora of new phenomena such as the quantum anomalous Hall~\cite{ref1} and quantum spin Hall effects~\cite{ref2}, and new materials, including, but not limited to,  Weyl semimetals~\cite{ref3}, topological superconductors~\cite{ref4} and  topological insulators~\cite{ref5}. One of the most fascinating aspects of topology is the bulk-boundary correspondence~\cite{ref6}, in which bulk system properties predict the existence of topologically protected modes localized at the boundary between topologically trivial and nontrivial systems. Due to their robustness with respect to environmental perturbations, topologically protected gapless edge excitations have been proposed as platforms for a wide range of potential applications, spanning from electric circuits to quantum computation~\cite{ref7}. Topological phases of matter have been considered in systems other than electronic ones such as ultracold atoms~\cite{coldatoms}, photonic crystals \cite{wang2009}, and mechanical systems \cite{huber2016}. \\
\begin{figure}[b!]
\centering
\includegraphics[width=1\linewidth]{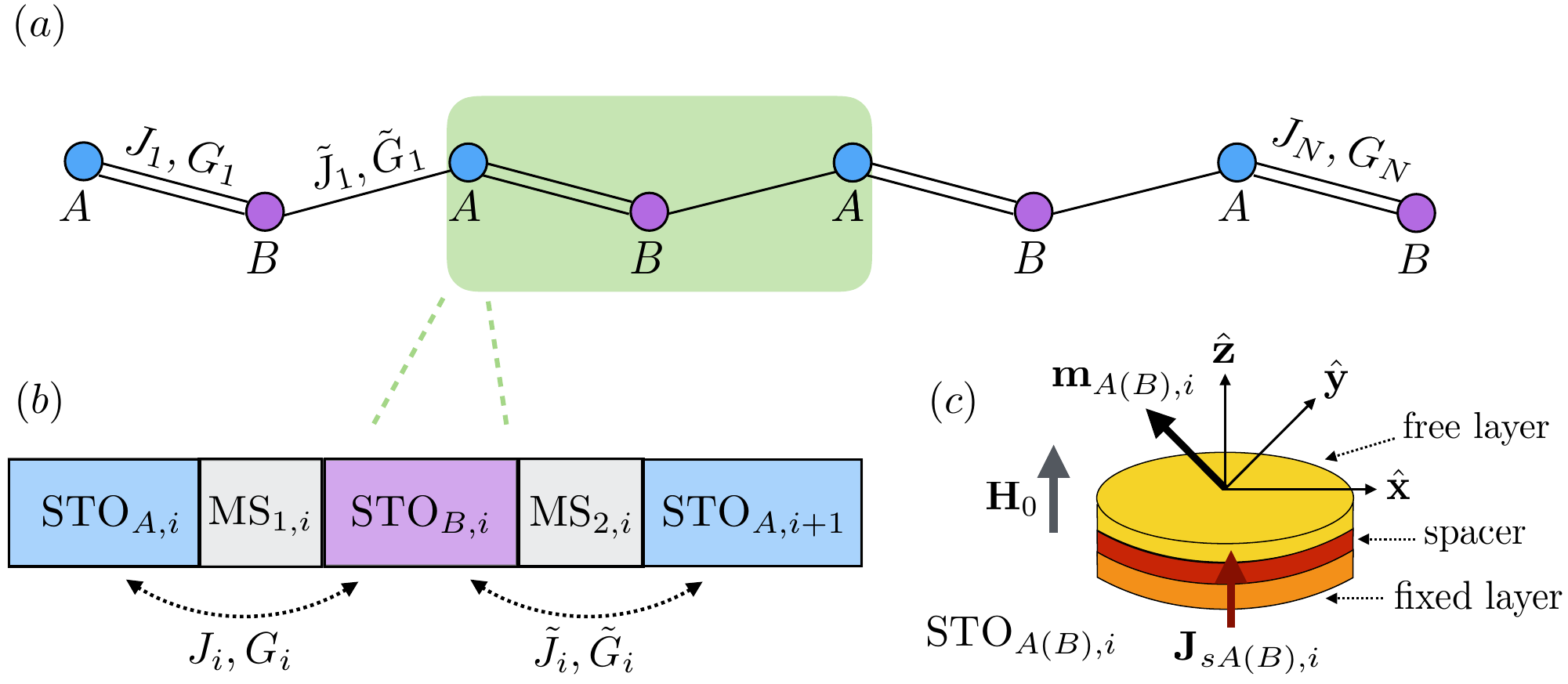}
\caption{Schematic illustration of our model. (a) A spin-torque oscillator (STO) array represented as a one-dimensional lattice with sublattices $A$ and $B$. Nearest-neighbor hopping in the $i$th unit cell (with $i=1,..., N$) is denoted $J_{i}, G_{i}$ and  $\tilde{J}_{i}, \tilde{G}_{i}$ for intra- and inter-cell hopping, respectively. (b) Each unit cell consists of two STOs, labelled $\text{STO}_{A,i}$ and $\text{STO}_{B,i}$, and two metallic spacers connecting adjacent STOs, denoted $\text{MS}_{1(2),i}$. Metallic spacers $\text{MS}_{1(2),i}$ mediate both a reactive  RKKY-like coupling and a dissipative spin-pumping coupling between spins of $\text{STO}_{A, i}$($\text{STO}_{B, i}$) and $\text{STO}_{B,i}$($\text{STO}_{A, i+1}$), whose strength is parametrized, respectively, by $J_{i}$($\tilde{J}_{i}$) and $G_{i}$($\tilde{G}_{i}$). (c) Each $\text{STO}$ is a free ferromagnet$|$spacer$|$fixed ferromagnet trilayer subjected to the spin current $\mathbf{J}_{sA(B),i}$. An external magnetic field $\mathbf{H}_{0}$ sets the equilibrium direction of the magnetic order parameter $\mathbf{m}_{A(B), i}$ of the free layer. Other implementations of STOs are possible.}.
\label{Fig1}
\end{figure}
\indent Topological bands for magnons, i.e. linear excitations of magnetic systems, have been considered \cite{topomagnons}, however this direction of research remains relatively unexplored. This is in part because magnon number is not conserved due to ubiquitous spin nonconserving interactions of magnons with the crystalline lattice~\cite{magnonphonon}. Conventional descriptions of topological phases rely on Hermitian Hamiltonians, which describe systems where the particle number is conserved.

The recent introduction of topological classifications for non-Hermitian Hamiltonians opens up new prospects for realizing nontrivial topological phases in intrinsically dissipative systems~\cite{ref8, Gong2018}. While a complete theoretical background it is still missing, recent experimental developments have shown that topologically protected edge modes can appear even when number conservation is violated~\cite{ref9}. 

In this Letter, we propose a non-Hermitian topological magnonic system. We consider a one-dimensional (1D) array of spin-torque oscillators (STOs)~\cite{STO}, which are current-driven magnetic nanopillars, as illustrated  in Fig.~\ref{Fig1}. We show that, in the linear regime, such an array is a magnonic realization of a non-Hermitian Su-Schrieffer-Heeger (SSH) model with Parity-Time (PT) symmetry or chiral-inversion (CI) symmetry~\cite{SSH,Lieu, song}, where different symmetry classes are accessed by changing experimentally controllable parameters. The crucial ingredients of our proposed setup are the injection of spin current into the magnets, which counteracts and may even overcome damping, and dissipative coupling of STOs via spin pumping. Using exact diagonalization and numerical solutions of the nonlinear equations describing magnetization dynamics, we find that this system displays a robust topological phase where a single STO at the edge is driven into auto-oscillation while the bulk remains inactive.

\textit{Model.} We consider a one-dimensional array of $2N$ STOs arranged in $N$ unit cells, as shown in Fig.~\ref{Fig1}. The  $i$th unit cell consists of two STOs,  labelled $\text{STO}_{A,i}$ and $\text{STO}_{B,i}$, and two metallic spacers coupling the intracell and intercell STO elements, denoted $\text{MS}_{1,i}$  and $\text{MS}_{2,i}$, respectively. The dynamics of the magnetic order parameter $\mathbf{m}_{\eta,i}$ (with $\eta=A,B$) of an isolated spin-torque oscillator  $\text{STO}_{\eta,i}$ subject to a magnetic field $\mathbf{H}_{0}=H_{0} \hat{\mathbf{z}}$ and a spin current (in units of frequency) $\mathbf{J}_{s \eta,i}= J_{s \eta,i} \hat{\mathbf{z}}$ is~\cite{ref11}
\begin{align}
\left. \dot{ \mathbf{m}}_{\eta,i} \right|_{0} &= \omega_{\eta,i} \hat{\mathbf{z}} \times \mathbf{m}_{\eta,i} + \alpha_{\eta,i}  \mathbf{m}_{\eta,i} \times  \dot{\mathbf{m}}_{\eta,i}    \nonumber \\ 
 &+ J_{s \eta,i} \mathbf{m}_{\eta,i} \times \left( \mathbf{m}_{\eta,i} \times \hat{\mathbf{z}}\right) \,.
\label{eq1}
\end{align}  

Here, $\omega_{\eta,i}=\gamma_{\eta,i} (H_{0}-4\pi M_{\eta,i})$ is the ferromagnetic resonance frequency, with $\gamma_{\eta,i}$ the gyromagnetic ratio, $M_{\eta,i}$  the saturation magnetization, and $\alpha_{\eta,i} \ll 1$ is the (dimensionless) Gilbert damping parameter. The second and third terms on the right-hand side of Eq.~(\ref{eq1}) are, respectively, the dissipative torque accounting for energy dissipation~\cite{gilbert} and the Slonczewski-Berger spin-transfer torque describing interaction of the magnetic order parameter with the spin-polarized current~\cite{torque}. In our proposed set-up, the spin transfer torque acts as a dissipative process that counteracts the intrinsic dissipation and thus provides tuneable gain. Metallic spacers can mediate both reactive and dissipative coupling between STOs. For the reactive coupling, we consider a RKKY-type exchange whose strength is parametrized by the frequency $J_{i}$  ($\tilde{J}_{i}$) for the spacer $\text{MS}_{1,i}$ ($\text{MS}_{2,i}$).  In our convention, $J_{i} (\tilde{J}_{i})>0$ corresponds to a ferromagnetic exchange coupling. The dissipative coupling is mediated by spin pumping through the spacer $\text{MS}_{1,i}$  ($\text{MS}_{2,i}$)~\cite{spinpumping}; its efficiency is parametrized by the dimensionless parameter $G_{i} (\tilde{G_{i}}) \ll 1$, which is microscopically related to the spin-mixing conductance of the oscillator$|$metallic spacer interface~\cite{spinmixing}. Assuming a nearest-neighbor coupling, the coupled dynamics introduced via the metallic spacers reads as
\begin{align}
\left. \dot{\mathbf{m}}_{A,i} \right|_{\rm coup}=&-\mathbf{m}_{A,i} \times ( J_{i} \mathbf{m}_{B,i} + \tilde{J}_{i-1} \mathbf{m}_{B,i-1}) \nonumber \\
 +&G_{i} \left[ \mathbf{m}_{A,i} \times \dot{\mathbf{m}}_{A,i} -\mathbf{m}_{B,i} \times \dot{\mathbf{m}}_{B,i} \right] \nonumber \\
 +& \tilde{G}_{i-1} \left[ \mathbf{m}_{A,i} \times \dot{\mathbf{m}}_{A,i} -\mathbf{m}_{B,i-1} \times \dot{\mathbf{m}}_{B,i-1} \right] \,, \nonumber \\
\left. \dot{\mathbf{m}}_{B,i} \right|_{\rm coup}=&-\mathbf{m}_{B,i} \times ( J_{i} \mathbf{m}_{A,i} + \tilde{J}_{i} \mathbf{m}_{A,i+1}) \nonumber \\
 +& G_{i} \left[ \mathbf{m}_{B,i} \times \dot{\mathbf{m}}_{B,i} -\mathbf{m}_{A,i} \times \dot{\mathbf{m}}_{A,i} \right] \nonumber \\
 +& \tilde{G}_{i} \left[ \mathbf{m}_{B,i} \times \dot{\mathbf{m}}_{B,i} -\mathbf{m}_{A,i+1} \times \dot{\mathbf{m}}_{A,i+1} \right]\,,
\label{eq2}
\end{align}
where  $\tilde{G}_{1(N)}=\tilde{J}_{1(N)}=0$. The sum of Eqs.~(\ref{eq1})~and~(\ref{eq2}) determines the full system dynamics. In the following, we consider identical unit cells and drop the parameter dependence on index $i$. We assume that the magnetic field is large enough to order the magnets along its direction, and proceed to linearize Eqs.~(\ref{eq1}) and~(\ref{eq2}) around the equilibrium direction of the magnetic order parameter. That is, we write $\mathbf{m}_{\eta,i}=(m_{\eta, i x}, m_{ \eta, i y}, 1),$ with $ |\mathbf{m}_{\eta, i}| \simeq 1$. Next, we introduce the complex variable $2 m^{-}_{\eta, i}=m_{\eta, i x}-im_{\eta,  i y}$ and invoke the Holstein-Primakoff transformation~\cite{Holstein} $m^{-}_{A(B),i}(t) =  \langle a_{i} (b_{i}) \rangle e^{-i\omega t} $, where the second-quantized operator $a_{i}$($b_{i}$) annihilates a magnon at the sublattice $A$($B$) of the $i$-th unit cell and obeys bosonic commutation relations. 

\begin{figure*}[t!]
\centering
\includegraphics[scale=1.0]{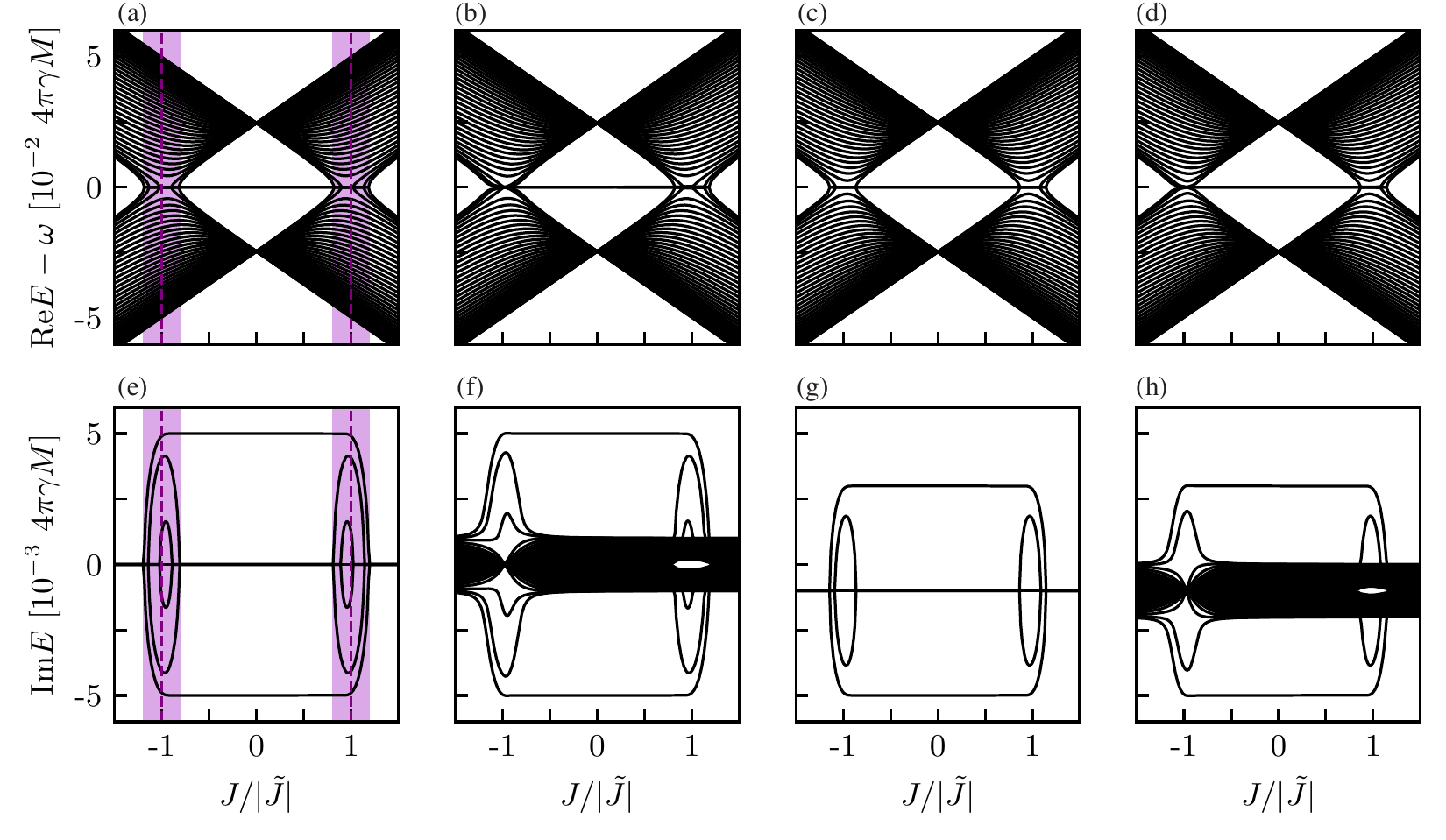}
\caption{ 
Exact diagonalization of Eqs.~\eqref{eq3}-\eqref{eq4} for $N=40$ unit cells ($80$ STOs) with $\omega/4\pi\gamma M = 0.5$, $\alpha = 0.01$, $\tilde{J}/4\pi\gamma M = 0.025$. The columns correspond to the four symmetry regimes of $H_k$ discussed in the main text. (a)-(d) Shows Re$E-\omega$ and (e)-(f) Im$E$. All cases have a line gap on the real axis and two edge modes with $\mathrm{Re} E-\omega = 0$ for $J \lesssim |\tilde{J}|$. In general, the real spectrum is mostly unaltered in each case, aside from small deviations of the  gap closing point. (a, e) PT symmetric case. There is one lasing edge mode for $J < |\tilde{J}-\alpha\omega|$, whereas additional bulk lasing modes appear for $|\tilde{J}-\alpha\omega| < J < |\tilde{J} + \alpha\omega|$ in the `PT-broken' regime~\cite{Lieu} (purple/light gray shaded area). The topological transition occurs at $J = \pm\tilde{J}$ (purple/gray dashed line). (b, f) Nonzero dissipative couplings $G, \tilde{G} = 0.1\alpha$. Bulk modes have Im$E\neq 0$, but there is still a well-defined lasing edge mode separated from the bulk modes for $J \lesssim |\tilde{J}|$. This regime has chiral-inversion (CI) symmetry. (c, g) Deviation of $J_{s}$ from the PT symmetric value $J_s^* = 2\alpha\omega$ where $J_s = 0.8J_{s}^*$. For $J_{s} < J_{s}^*$ the imaginary spectrum is shifted down, while for $J_{s} > J_{s}^*$ it is shifted up (not shown). The edge mode is still lasing ($\mathrm{Im} E >0$) for $J_{s} \gtrsim \alpha\omega$. This regime has chiral symmetry (CS). (d, h) Combination of the two PT breaking terms, $G, \tilde{G} = 0.1\alpha$ and $J_s=0.8J_s^*$. Im$E$ has properties of (f) and (g). Both CI and CS are broken, but TRS$^\dagger$ symmetry is preserved.}\label{Fig:bandstructure}
\end{figure*}

Assuming the local damping to be much larger than the nonlocal one, $\tilde{G}, G \ll \alpha_{\eta}$, the linearized dynamics that follows from Eqs.~\eqref{eq1}-\eqref{eq2} corresponds to a non-Hermitian SSH model with complex onsite potentials and complex intercell and intracell couplings. For identical STOs on the $A$ and $B$ sublattices we have $\omega_{\eta}\equiv \omega$ and $\alpha_{\eta}\equiv \alpha$.  The corresponding Hamiltonian is $H=\sum_{i=1}^{N} H_{i}$, with $H_{i}$ the Hamiltonian for the $i$th unit cell, which reads
\begin{align}
H_{i} &=  \omega \left[ a^{\dagger}_{i}  a_{i} +  b^{\dagger}_{i}  b_{i} \right] +i  (J_{sA}-\alpha  \omega) a^{\dagger}_{i}  a_{i}\nonumber \\
&+ i   (J_{sB}-\alpha \omega)  b^{\dagger}_{i}  b_{i} + (-J+i G \omega) \left[a^{\dagger}_{i} b_{i} + \mathrm{h.c.}\right]\nonumber \\
&+ (-\tilde{J}+i  \tilde{G} \omega) \left[ a^{\dagger}_{i} b_{i-1} + \mathrm{h.c.}\right]~,
\label{eq3}
\end{align}
for $i\neq 1 , N$, with open boundary conditions
 \begin{align}
 H_{j}&= \omega a^{\dagger}_{j}  a_{j}+\omega b^{\dagger}_{j} b_{j} +i (J_{sA}-\alpha \omega) a^{\dagger}_{j}  a_{j} 
\nonumber \\ &+i (J_{sB}-\alpha \omega ) b^{\dagger}_{j}  b_{j}  +  (-J+iG \omega) \left[ a^{\dagger}_{j} b_{j} + b^{\dagger}_{j} a_{j}\right] \nonumber \\ & + (-\tilde{J} + i \tilde{G} \omega) b^{\dagger}_{j} a_{l}~, 
\label{eq4}
\end{align}
with $j=1,N$ and $l=2, N-1$.
Equations~\eqref{eq3}-\eqref{eq4} constitute our starting point for exact diagonalization. 

\textit{Results.}~The effective Hamiltonian in Eqs.~\eqref{eq3}-\eqref{eq4} is a realization of a non-Hermitian SSH chain with many tunable features affecting its symmetries.  We investigate the topology as a function of $J/\tilde{J}$, which can be experimentally controlled by tuning the length of the metallic spacers. Here, we consider the case where spin current is injected only at the $A$ sites, setting $J_{sA} = J_s$ and $J_{sB} = 0$. Importantly, $H$ does not exhibit the non-Hermitian skin effect and therefore analysis of its topology based on periodic boundary conditions is valid~\cite{song, kunst2018, ref8}. The Fourier transform of Eq.~\eqref{eq3} for periodic boundary conditions is  
\begin{equation}
H_k = id_{0k}\mathbb{I} + \vec{d}_k \cdot \boldsymbol{\sigma}, 
\label{Eqn:Hkgeneral}
\end{equation}
where $\mathbb{I}$ is the $2\times2$ identity matrix, $\boldsymbol{\sigma} = (\sigma^x, \sigma^y, \sigma^z)$ is the vector of Pauli matrices, $\vec{d}_k = (d_{xk}, d_{yk}, d_{zk}) \in \mathbb{C}$ is a vector of complex functions of momentum $k$ and $d_{0k} \in \mathbb{R}$ is a real function of $k$. The energy spectrum is 
\begin{equation}
E^\pm_k= id_{0k} \pm \sqrt{\vec{d}_{1k}^2 - \vec{d}_{2k}^2 + 2i\vec{d}_{1k}\cdot\vec{d}_{2k}},
\label{Eqn:energygeneral}
\end{equation}
where $\vec{d}_{1k} = \mathrm{Re}\left[\vec{d}_k\right]$, $\vec{d}_{2k} = \mathrm{Im}\left[\vec{d}_k\right]$. Here, $d_{0k} = (J_s-2\alpha\omega)/2$ and 
\begin{equation}
\vec{d}_k = \begin{pmatrix}
-J +i\omega G -(\tilde{J} - i\tilde{G}\omega)\cos k \\
-(\tilde{J}-i\tilde{G}\omega)\sin k \\
i\frac{J_s}{2}
\end{pmatrix}.
\end{equation}
$E^\pm_k$ has a real line gap where $\mathrm{Re}[E^\pm_k] \neq 0~\forall~k$ provided that $\vec{d}_{1k}^2 > \vec{d}_{2k}^2~\forall~k$. Systems with a real line gap can have topogically nontrivial phases depending on their symmetry class~\cite{Gong2018, ref8, kunst2018}. Furthermore, since this model includes only short-range hopping, the condition for a topological phase transition can be formulated as in Ref.~\cite{kunst2018}. We use the symmetry naming conventions of Ref.~\cite{ref8}. We consider four different cases, corresponding to the {\it i)} PT symmetric case, {\it ii,iii)} two different types of breaking of PT symmetry and {\it iv)} a combination of both symmetry-breaking terms.\\
\indent \emph{i)} For $J_{s} = J_s^* = 2\alpha\omega$ and negligible dissipative couplings, $G, \tilde{G}=0$, $H_k$ is PT symmetric with $\sigma_{x} H^*_{k} \sigma_{x}=H_{k}$. This system is a well-known host of two topologically protected edge modes with $\mathrm{Im}E =\pm 2\alpha \omega$  for $|J| < |\tilde{J}|$, and it has a real line gap for $|J-\tilde{J}| > \alpha\omega$~\cite{Lieu}. The mode with positive  (negative) imaginary energy, i.e., the lasing (lossy) edge mode, corresponds physically to a magnon population that grows (decays) exponentially in time at the left (right) edge. The PT-symmetric model has generated much interest and has been experimentally realized in photonic systems and microresonators~\cite{ref8, ref9}. \\
\indent \emph{ii)} For $J_s = J_s^*$ and $G(\tilde{G}) \neq 0$, Eq.~\eqref{Eqn:Hkgeneral} obeys chiral-inversion (CI) symmetry $\sigma_y H_k\sigma_y = -H_{-k}$. CI symmetric models have nontrivial topological phases~\cite{song}. The bulk real gap is open for $|J-\tilde{J}| > \omega\sqrt{\delta G^2 +\alpha^2}$,  with $\delta G = G-\tilde{G}$, and the condition for two topologically protected edge modes 
is modified to $| J-iG\omega| < |\tilde{J}-i\tilde{G}\omega|$.\\
\indent \emph{iii)} Deviations of $J_s$ from $J_s^*$ with $G, \tilde{G} = 0$ preserve chiral symmetry (CS) defined as $\sigma_z H^\dagger \sigma_z = -H_k$. In this case there is a real gap for $|J-\tilde{J}| > J_s/2$ and exhibits a topological phase with two protected edge modes for $|J| < |\tilde{J}|$. \\
\indent \emph{iv)} Including both $G(\tilde{G}) \neq 0$ and $J_s \neq J_s^*$ breaks CI and CS symmetries. The Hamiltonian maintains TRS$^\dagger$ symmetry defined as $H^T(k) = H(-k)$ and has a real gap for $|J-\tilde{J}| > \sqrt{\delta G^2\omega^2 +(J_s/2)^2}$; however, TRS$^\dagger$ alone does not guarantee topologically protected phases in 1D~\cite{ref8}. \\
\indent This analysis shows that with dissipative coupling \emph{or} injected spin current $J_s \neq J_s^*$ the model still has a topologically nontrivial phase.  Furthermore, only the difference $\delta G$ affects the topological transition and bulk gap closing. Thus, for the parameter regime $G, \tilde{G} \ll \alpha$ and $G \approx \tilde{G}$, in practice resulting from choosing the metallic spacers to be good spin sinks, the edge modes are quite robust. Including both PT symmetry breaking terms \emph{may} compromise the topological protection as PT, CS, and CI symmetries are all broken. We find numerically that localized edge modes are still present for $J_s \neq J_s^*$ and $G, \tilde{G} = 0.1\alpha$, but these modes may not be topologically protected~\cite{ref8}.\\
\indent The complex frequency spectrum $E$ that results from exact diagonalization of Eqs.~\eqref{eq3}-\eqref{eq4} is presented in Fig.~\ref{Fig:bandstructure} for all cases discussed above. For weak dissipative coupling $G, \tilde{G}\ll \alpha$ and $\sim 20\%$ deviations of $J_s$ from the PT symmetric value $J_s^*$ we always find a real line gap for $|J| \lesssim |\tilde{J}|$, shown in Fig.~\ref{Fig:bandstructure}~(a)-(d), and a lasing edge mode with $\mathrm{Im}E >0$ shown in Fig.~\ref{Fig:bandstructure}~(e)-(h). The lasing mode is well separated from the imaginary spectrum of bulk states.\\
\begin{figure}[t!]
\centering
\includegraphics[width=\columnwidth]{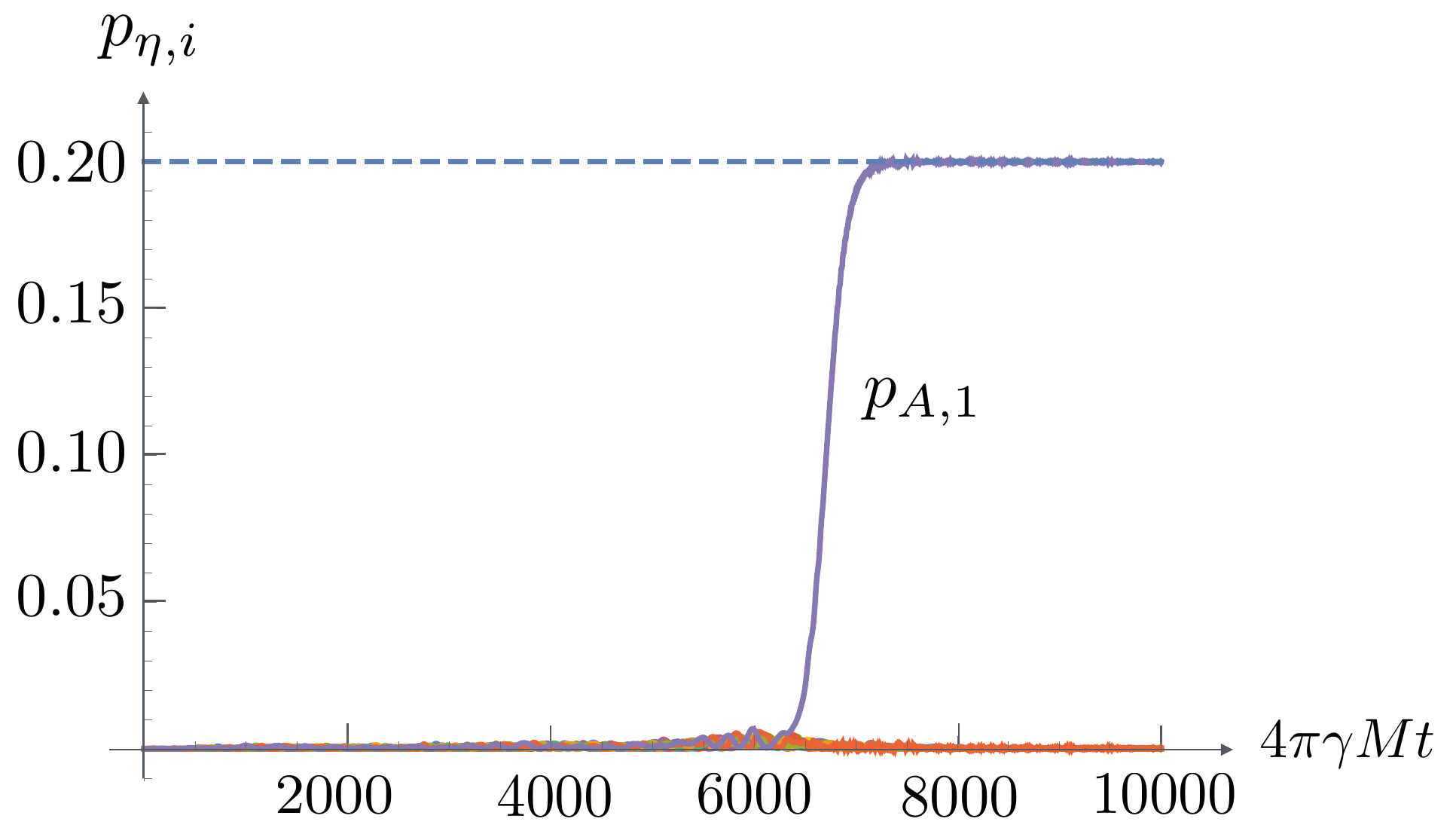}
\caption{Power $p_ {\eta, i}$ of the oscillator with $\mathbf{m}_{\eta,i}$ in an array of $N=10$ unit cells as a function of time (normalized by the frequency $4\pi\gamma M$). The dashed line corresponds to the steady-state power of a single oscillator. Parameters taken are $2 k_B T/N_s \hbar \omega=10^{-4}$, $\alpha = 10^{-2}$, $\tilde J = -0.025/4\pi\gamma M$, $\omega/4\pi\gamma M=0.5$, $J/\tilde J=0.5$, and $J_s=J_s^*$.  }\label{Fig:lasingSTO}
\end{figure}
\emph{Numerical solutions}.~To investigate the nonlinear dynamics of the STO array, and, in particular, how nonlinearities affect the growth of the lasing edge modes, we solve Eqs.~\eqref{eq1}-\eqref{eq2} numerically. We parametrize the magnetization direction by the power $0< p_{\eta,i}(t) <1$, proportional to the experimentally emitted microwave power, and a precession angle $\phi_{\eta,i} (t)$ as
\begin{align}
\mathbf{m}_{\eta,i}= \begin{pmatrix} 2 \sqrt{p_{\eta,i} (1-p_{\eta,i})} \sin \phi_{\eta,i} \\ 2 \sqrt{p_{\eta,i} (1-p_{\eta,i})} \cos \phi_{\eta,i} \\ 1 -2p_{\eta,i} \end{pmatrix}\,.
\end{align} 
The initial value of $\phi_{\eta,i} (t=0)$ is taken to be a random number between $0$ and $2\pi$ and the initial power $p_{\eta,i} (t=0)$ is drawn from a thermal equilibrium distribution for an ensemble of isolated STOs~\cite{ref11}, 
\begin{equation}
P_{\rm eq} \propto \exp \left[-\frac{N_{s}}{ k_B T} \omega  p_{\eta,i} \right]\,,
\end{equation}
where $k_{B}$ is the Boltzmann constant, $T$ the temperature, and $N_{s}$ the number of spins in each STO. We take $2 k_B T/N_s \hbar \omega=10^{-4}$ and $\alpha = 10^{-2}$ for typical materials used for spin-torque experiments at room temperature~\cite{ref11}. In Fig.~\ref{Fig:lasingSTO} we show the numerical results for the power of all oscillators for $\tilde J = -0.025/4\pi\gamma M$, $\omega/4\pi\gamma M=0.5$, $J/\tilde J=0.5$, and $J_s=J_s^*$, which corresponds to the PT symmetric case of one lasing edge mode. 

We find that, after some time, the power of the oscillator on the right edge grows and eventually saturates while the other oscillators have negligible power at all times. The power saturation level of the oscillator at the edge is, because of the small couplings $J$ and $\tilde J$, well approximated by the steady-state power of a single oscillator as found from Eq.~(\ref{eq1}) (see Ref.~\cite{ref11}), denoted by a dashed line in Fig.~\ref{Fig:lasingSTO}. Our numerical results therefore confirm the presence of a lasing edge mode, which manifests as a topologically protected auto-oscillation of the oscillator at the edge. While we find from our numerical solutions that the dynamics outside the regime of one topologically-protected lasing edge mode is interesting, it is also non-universal. For example, often several (but not all) oscillators reach a steady state at nonzero power, but which oscillators reach nonzero power strongly depends on the initial conditions. We consider an exhaustive study of the nonlinear dynamics of the proposed system beyond the scope of this article. 

\textit{Discussion.} In this work, we classified the topology of an intrinsically dissipative magnetic system. We establish a mapping between the linearized magnetic dynamics of a 1D array of STOs and a non-Hermitian SSH model. We find that topologically nontrivial phases can be accessed by tuning both the spin current injection and the properties of the metallic spacers connecting the STOs. The topological phases support a topologically protected lasing edge mode, which manifests as a single edge STO emitting a microwave signal while the bulk STOs are not active. Our results show that the topologically-protected edge mode is robust against deviations of the spin current from the PT symmetric value and robust against dissipative coupling between STOs.

The emergence and position of the lasing edge mode is controlled via spin current injection. The location of the lasing edge switches sides upon changing spin-current injection from $A$ to $B$ sublattice sites, thereby opening up prospects for building tunable and robust spin-wave waveguides and neuromorphic networks~\cite{ref10}. Future work should address the effects of long-range intercell and intracell couplings, the effects of drives and coupled dynamics~\cite{cavity}, and the non-Hermitian topology of magnetic systems in higher dimensions~\cite{yao2018}.

\textit{Acknowledgements.} This research was supported in part by the National Science Foundation under Grant No. NSF PHY-1748958. HMH acknowledges the support of an NRC Research Assistantship at NIST. This work is supported by the European Research Council via Consolidator Grant number
725509 SPINBEYOND. RD is member of the D-ITP consortium, a program of the Netherlands Organisation for
Scientific Research (NWO) that is funded by the Dutch
Ministry of Education, Culture and Science (OCW).


\begin{thebibliography}{99}


\bibitem{topo} K. v. Klitzing, G. Dorda, and M. Pepper, Phys. Rev. Lett. \textbf{45}, 494 (1980); D. J. Thouless, M. Kohmoto, M. P. Nightingale, and M. den Nijs, Phys. Rev. Lett. \textbf{49}, 405 (1982).
\bibitem{ref1} C.-X. Liu, S.-C. Zhang, and X.-L. Qi, Annu. Rev. Condens. Matter Phys. \textbf{7}, 301 (2016). 
\bibitem{ref2} J. Maciejko, T. L. Hughes, and S.-C. Zhang, Annu. Rev. Condens. Matter Phys. \textbf{2}, 31 (2011).
\bibitem{ref3} B. Yan and C. Felser, Annu. Rev. Condens. Matter Phys. \textbf{8}, 337 (2017).
\bibitem{ref4} M. Sato and Y. Ando, Rep. Prog. Phys. \textbf{80}, 7 (2017).
\bibitem{ref5}  J. E. Moore, Nature  \textbf{464}, 194 (2010).
\bibitem{ref6}  X.-L. Qi, H. Katsura,  and A. W. W. Ludwig,  Phys. Rev. Lett. \textbf{108}, 196402 (2012).
\bibitem{ref7}  P. Liu, J. R. Williams, and J. J. Cha, Nat. Rev. Mat. \textbf{4},  479 (2019).
\bibitem{coldatoms} B. K. Stuhl, H. I. Lu, L. M. Aycock, D. Genkina, and I. B. Spielman, Science \textbf{349} (6255), 1514 (2015); D.-W. Zhang, Y.-Q. Zhu, Y.X. Zhao, H. Yan, S.-L. Zhu, Adv. Phys. \textbf{67}, 253 (2019).
\bibitem{wang2009} Zheng Wang, Yidong Chong, J. D. Joannopoulos and Marin Soljacic, Nature {\bf 461}, 772 (2009).
\bibitem{huber2016} Sebastian D. Huber, Nat. Phys. {\bf 12}, 621 (2016).
\bibitem{topomagnons} R. Shindou, R. Matsumoto, S. Murakami, and J. Ohe, Phys. Rev. B {\bf 87}, 174427 (2013);
R. Chisnell, J. S. Helton, D. E. Freedman, D. K. Singh, R. I. Bewley, D. G. Nocera, and Y. S. Lee, Phys. Rev. Lett. {\bf 115}, 147201 (2015); S. K. Kim, H. Ochoa, R. Zarzuela, and Y. Tserkovnyak, Phys. Rev. Lett. {\bf 117}, 227201 (2016); Andreas R\"uckriegel, Arne Brataas, and R.A. Duine, Phys. Rev. B \textbf{97}, 081106 (2018);
Lebing Chen, Jae-Ho Chung, Bin Gao, Tong Chen, Matthew B. Stone, Alexander I. Kolesnikov, Qingzhen Huang, and Pengcheng Dai, Phys. Rev. X {\bf 8}, 041028 (2018).
\bibitem{magnonphonon}  T. Kikkawa, K. Shen, B. Flebus, R. A. Duine, K. Uchida, Z. Qiu, G. E. W. Bauer, and E. Saitoh,  Phys. Rev. Lett. \textbf{117}, 207203 (2016); B. Flebus, K. Shen, T. Kikkawa, K. Uchida, Z. Qiu, E. Saitoh, R. A. Duine, and G. E. W. Bauer,  Phys. Rev. B \textbf{95}, 144420 (2017); S. Streib, N. Vidal-Silva, K. Shen, and G. E. W. Bauer
Phys. Rev. B \textbf{99}, 184442.
\bibitem{ref8} K. Kawabata, K. Shiozaki, M. Ueda, and M. Sato, Phys. Rev. X \textbf{9}, 041015 (2019).
\bibitem{Gong2018} Z. Gong, Y. Ashida, K. Kawabata, K. Takasan, S. Higashikawa, and M. Ueda,  Phys. Rev. X \textbf{8}, 031079 (2018); M. A. Bandres and M. Segev,  Physics \textbf{11}, 96 (2018).
\bibitem{ref9} S. Weimann, M. Kremer, Y. Plotnik, Y. Lumer, S. Nolte, K. G. Makris, M. Segev, M. C. Rechtsman, and A. Szameit,  Nat. Mater. \textbf{16}, 433 (2017);  M. Pan, H. Zhao, P. Miao, S. Longhi, and L. Feng,  Nat. Commun. \textbf{9}, 1308 (2018); M. Parto, S. Wittek, H. Hodaei, G. Harari, M. A. Bandres, J. Ren, M. C. Rechtsman, M. Segev, D. N. Christodoulides,  and M. Khajavikhan, Phys. Rev. Lett. \textbf{120}, 113901 (2018); R. El-Ganainy, M. Khajavikhan, D. N. Christodoulides, and S. K. Ozdemir, Nat. Commun. Phys. \textbf{2}, 37 (2019); C. Poli, M. Bellec, U. Kuhl, F. Mortessagne, and
H. Schomerus, Nat. Comm. \textbf{6}, 6710 (2015);  J. M. Zeuner, M. C. Rechtsman, Y. Plotnik, Y. Lumer, S. Nolte, M. S. Rudner, M. Segev, and A. Szameit, Phys.
Rev. Lett. \textbf{115}, 040402 (2015); L. Xiao, X. Zhan, Z. H. Bian, K. K. Wang, X. Zhang,
X. P. Wang, J. Li, K. Mochizuki, D. Kim, N. Kawakami, W. Yi, H. Obuse, B. C. Sanders, P. Xue,  Nat. Phys. \textbf{13}, 1117 (2017).


\bibitem{STO} T. Chen, R. K. Dumas, A. Eklund, P. K. Muduli, A. Houshang, A. A. Awad, P. D$\ddot{\text{u}}$rrenfeld,
B. G. Malm, A. Rusu, and Johan $\dot{\text{A}}$kerman, Proc. of the IEEE \textbf{104}, 10 (2016); J.-V. Kim,  Solid State Physics \textbf{63}, 217 (2012).
\bibitem{SSH} W. P. Su, J. R. Schrieffer, and A. J. Heeger, Phys. Rev. Lett. \textbf{42}, 1698 (1979).
\bibitem{Lieu} K. Esaki, M. Sato, K. Hasebe, and M. Kohmoto,
Phys. Rev. B \textbf{84}, 205128 (2011); B. Zhu, R. L$\ddot{\text{u}}$, and S. Chen,
Phys. Rev. A \textbf{89}, 062102, (2014); S. Lieu, Phys. Rev. B \textbf{97}, 045106 (2018).
\bibitem{song} L. Jin and Z. Song, Phys. Rev. B \textbf{99}, 081103 (2019);

\bibitem{ref11} A. Slavin and V. Tiberkevich, IEEE Trans. Magn. \textbf{45}, 1875 (2009).
\bibitem{gilbert} T. L. Gilbert, IEEE Trans. Magn. \textbf{40}, 3443 (2004).
\bibitem{torque} J. C. Slonczewski,  J. Magn. Magn. Mater. \textbf{159},  L1 (1996); L. Berger, Phys. Rev. B \textbf{54},  9353 (1996).
\bibitem{spinpumping} B. Heinrich, Y. Tserkovnyak, G. Woltersdorf, A. Brataas, R. Urban, and G. E. W. Bauer, Phys. Rev. Lett.
\textbf{90}, 187601 (2003); Y. Tserkovnyak, A. Brataas, and
G. E. W. Bauer, Phys. Rev. B \textbf{67}, 140404(R) (2003).
\bibitem{spinmixing}  Y. Tserkovnyak, A. Brataas, G. E. W. Bauer, and B. I. Halperin, Rev. Mod. Phys. \textbf{77}, 1375 (2005).
\bibitem{Holstein} T. Holstein, and H. Primakoff, Phys. Rev. \textbf{58}, 1098 (1940).
\bibitem{kunst2018} F. K. Kunst, E. Edvardsson, J. C. Budich, and E. J. Bergholtz, Phys. Rev. Lett. \textbf{121}(2), 026808 (2018). 
\bibitem{ref10} N. Locatelli, V. Cros, and  J. Grollier,  Nat. Mat. \textbf{13},  11 (2014).
\bibitem{cavity} Y. Cao and P. Yan, Phys. Rev. B \textbf{99}, 214415 (2019).
\bibitem{yao2018} H. Shen, B. Zhen, and L. Fu
Phys. Rev. Lett. \textbf{120}, 146402 (2018); Y. Shunyu, and Z. Wang. Phys. Rev. Lett. \textbf{121}(8), 086803 (2018).


\end{thebibliography}
\end{document}